\documentclass[aps,prl,twocolumn,amsmath,amssymb,amsfonts,nofootinbib,long,floatfix,showpacs]{revtex4}
\usepackage{epsfig,latexsym,bm}

\newcommand\GeV{\mbox{GeV}}

\newcommand\G{\mbox{G}}
\newcommand\A{\mathbf{A}}
\newcommand\B{\mathbf{B}}
\newcommand\x{\mathbf{x}}
\newcommand\kk{\mathbf{k}}
\newcommand\ee{{\boldsymbol \varepsilon}}
\newcommand\E{\mathbf{E}}

\begin{document}

\title{Cosmic magnetization out from the vacuum}

\author{Leonardo Campanelli$^{1}$}
\email{leonardo.campanelli@ba.infn.it}
\affiliation{$^1$Dipartimento di Fisica, Universit\`{a} di Bari, I-70126 Bari, Italy}

\date{\today}


\begin{abstract}
The large-scale magnetic fields we observe today in galaxies and galaxy clusters could be the
result of a pure quantum effect taking place during inflation,
to wit, the creation of particles (photons) out from the vacuum in a curved spacetime.
We show that, whenever the conformal invariance of electromagnetism is broken during inflation,
the actual magnetic field spectrum, in the classical limit, is given by $B_k \simeq k^2 \sqrt{n_\kk}$,
where $n_\kk \gg 1$ is the number of created photons with wavenumber $k$.
In particular, a scale-invariant magnetic field of order of $10^{12} \G$ can emerge in the
simplest model of cosmic magnetogenesis, the one where the inflaton
is kinetically coupled to the photon.
Moreover, and contrarily to the general belief,
we show that such a model is free from the so-called strong-coupling and
backreaction problems.
This conclusion follows, indirectly, from the observation that post-inflationary electric currents,
which in the literature are incorrectly supposed to freeze superhorizon magnetic fields after inflation,
are indeed vanishing on superhorizon scales due to causality.
\end{abstract}




\maketitle


{\it Introduction.} -- Long time ago, Turner and Widrow~\cite{Turner-Widrow} advanced the idea that
the large-scale magnetic fields we observe today in galaxies and galaxy clusters
(for a recent review on cosmic magnetic fields, see~\cite{Subramanian}) could be nothing else that
relics from inflation. They also realized that, since Maxwell theory is conformally invariant
in conformal-invariant spacetimes, classical magnetic fields (as those we observe today)
cannot be produced in standard electrodynamics. From this, the need arose to introduce nonstandard
conformal-breaking terms in the electromagnetic Lagrangian, such as nonminimal couplings between
electromagnetism and gravity. Some years later, Ratra~\cite{Ratra} proposed a model in which
the inflaton $\phi$ (the scalar field responsible for inflation) is kinematically coupled to the photon
via an exponential coupling of the form $e^{\phi} F^2$.
This model has been successful for two reasons. First, it is able to produce scale-invariant magnetic fields
whose intensity ($\sim 10^{-12} \G$) can directly explain cosmic magnetic fields.
Secondly, the kinematic coupling, which was just postulated by Ratra,
has been showed, by Martin and Yokoyama~\cite{Martin-Yokoyama}, to naturally emerge in string theory.
However, soon after the framing of the Ratra model in a well-motivated, high-energy particle physics model,
Demozzi, Mukhanov and Rubinstein~\cite{Demozzi}
pointed out that inflation-produced magnetic fields able to explain cosmic magnetization are
either in an unmanageable strong-coupling regime during inflation or strongly back-react on the dynamics
of the inflaton. This sort of  ``no-go theorem'' for inflationary magnetogenesis not only
invalidated the predictions of the very elegant and simple model proposed by Ratra, but also
the results of a plethora of other generating mechanisms put forward in the literature since
the appearance of the seminal paper by Turner and Widrow.

However, there is a flaw in the arguments of all the above cited works: it is
assumed that post-inflationary electric currents froze the inflation-produced
magnetic field on all scales, including superhorizon scales which are those of
astrophysical interest for cosmic magnetic fields. But this would imply a violation of
causality since, as first pointed out by Barrow and Tsagas~\cite{Barrow},
post-inflationary currents are generated by microphysical processes during reheating and are, then,
always vanishing on superhorizon scales.

The aim of this paper is to reconsider the generation of primordial magnetic fields at inflation
in the kinetically coupled scenario by studying
the creation of photons out from the vacuum in an expanding universe, a very peculiar phenomenon of
non-conformal-invariant quantum field theories in curved spacetime~\cite{Birrell-Davies}.
In this framework, we will fix the ``causality flaw'' and, at the same time, we will
show that there is no ``no-go theorem'' for inflationary magnetogenesis in the Ratra model,
which indeed fully accounts for cosmic magnetic fields.

{\it The model.} -- It can be shown that the most general Lorentz-, C-, P-, CP-,
CPT-, gauge-invariant, minimally coupled electromagnetic Lagrangian in four spacetime dimensions,
which involves classical, homogeneous, background fields,
has the simple form ${\mathcal L}_{\rm em} = f(\phi_i){\mathcal L}_{\rm M}$.
Here, ${\mathcal L}_{\rm M} = -\frac14 F^2$
is the standard (free) Maxwell Lagrangian, and the coupling function $f(\phi_i)$ incorporates
all possible interactions of the photon with scalar fields $\phi_i$, which can be either composite or,
as the inflaton $\phi$ in the Ratra model, elementary.
In order to have a positive-defined electromagnetic energy,
and to avoid the strong-coupling problem~\cite{Demozzi},
$f(\phi_i)$ must be a positive-defined quantity greater than unity.
Finally, in order to recover standard electromagnetism
and not to spoil, thus, the predictions of the standard cosmological model,
$f(\phi_i)$ must tend to one after reheating.

We restrict our analysis to the case of a spatially flat universe
described by the line element $ds^2 = a^2(d\eta^2 - d \x^2)$,
where $\eta$ is the conformal time and
$a(\eta)$ is the expansion parameter
normalized to unity at the present conformal time $\eta_0$.
Due to homogeneity and isotropy of both metric and background fields,
the coupling $f(\phi_i)$ turns to be a function only of the conformal time.

{\it Particle creation.} -- We now quantize our model.
A general characteristic of quantum fields in curved spacetime is that,
since Poincar\'{e} group is not in general a symmetry of the spacetime,
the vacuum state cannot be uniquely defined as in the case of
quantum theory in Minkowski spacetime. Nevertheless,
we can define an ``adiabatic'' in-vacuum state $|0, {\rm in} \rangle$
such that it is annihilated by the annihilation operators, $a_{\kk,\lambda}^{({\rm in})} |0, {\rm in} \rangle = 0$,
for all photon comoving wavenumbers $\kk$ and polarization $\lambda$, and normalized as $\langle 0, {\rm in}|0, {\rm in}\rangle = 1$.
The in-vacuum state contains no in-particles in the ``in-region'', $\eta \rightarrow -\infty$,
which will represent the temporal region when inflation starts.
Such a vacuum can be constructed by solving the equation of motion for the photon polarization states
and then by fixing the constants of integrations appearing in the general solution
by matching the latter with the corresponding adiabatic solution for $\eta \rightarrow -\infty$~\cite{Birrell-Davies}.
Accordingly, we can construct the in-Fock space based on $|0, {\rm in} \rangle$ by repeatedly applying the creation operator
$a_{\kk,\lambda}^{({\rm in}) \dag}$ on the in-vacuum state.
The electromagnetic vector potential in the in-Fock space reads
\begin{equation}
\label{A}
{\A}(\eta,\x) =
\sum_{\lambda=1}^2 \int \!\! \frac{d^3k}{(2\pi)^3} \frac{\ee_{\kk,\lambda}}{ \sqrt{2k}} \,
a_{\kk,\lambda}^{({\rm in})} \, A_{k,\lambda}^{({\rm in})} \, e^{i\kk \x} + \mbox{H.c.},
\end{equation}
where $\ee_{\kk,\lambda}$ are the standard circular polarization vectors.
The vector potential and its conjugate momentum
$\boldsymbol{\pi} = f \dot{\A}$,
as well as the annihilation and creation operators,
satisfy the usual commutation relations (hereafter, a dot indicates differentiation with respect to the conformal time).
The photon wave function $A_{k,\lambda}^{({\rm in})}(\eta)$ is a solution of the equation of motion
$\ddot{A}_{k,\lambda} - (\dot{f}/f) \dot{A}_{k,\lambda} + k^2 A_{k,\lambda} = 0$,
and it matches the corresponding adiabatic solution in the in-region.
In order to have a self-consistent quantization, it must satisfy the Wronskian condition
$A_{k,\lambda}^{({\rm in})} \dot{A}_{k,\lambda}^{({\rm in})*} - \dot{A}_{k,\lambda}^{({\rm in})} A_{k,\lambda}^{({\rm in})*} = 2ik/f$.

Consider, now, an ``out-region'' defined by $\eta \rightarrow +\infty$,
which will represent any temporal region after reheating.
An observer in this region may define an out-vacuum state
$|0, {\rm out} \rangle$ and construct from it the appropriate out-Fock space.
He can then expand the electromagnetic field
in terms of the out-annihilation and out-creation operators,
with the photon wave function $A_{k,\lambda}^{({\rm out})}(\eta)$ matching the corresponding
adiabatic solution at $\eta \rightarrow +\infty$.

It is clear from the above discussion that the two Fock spaces based on the two different
choices of the vacuum state are both physically
acceptable and must be then related. In particular, there will be a relation
between the in- and out-modes $A_{k,\lambda}^{({\rm in})}$ and $A_{k,\lambda}^{({\rm out})}$,
as well as a relation between the in- and out-creation and annihilation operators.
In order to find these relations, let us re-scale the electromagnetic field as
$\psi_{k,\lambda}^{({\rm in,out})} = \sqrt{f/2k} \, A_{k,\lambda}^{({\rm in,out})}$.
The in- and out-$\psi$ modes satisfy the equation of motion
$\ddot{\psi}_{k,\lambda} = U_k {\psi}_{k,\lambda}$,
with $U_k(\eta) = -k^2 + \; \ddot{\!\!\!\!\sqrt{f}}/\sqrt{f}$.
If $\psi_{k,\lambda}^{(1)}$ and $\psi_{k,\lambda}^{(2)}$ are two solutions of this equation,
the following inner product is conserved,
$\langle \psi_{k,\lambda}^{(1)} | \psi_{k,\lambda}^{(2)} \rangle =
-i (\psi_{k,\lambda}^{(1)} \dot{\psi}_{k,\lambda}^{(2)} - \dot{\psi}_{k,\lambda}^{(1)} \psi_{k,\lambda}^{(2)})$.
We can then introduce the time-independent quantities
$\alpha_{k,\lambda} = \langle \psi_{k,\lambda}^{({\rm in})} | \psi_{k,\lambda}^{({\rm out}) *} \rangle$
and $\beta_{k,\lambda} = -\langle \psi_{k,\lambda}^{({\rm in})} | \psi_{k,\lambda}^{({\rm out})} \rangle$.
Accordingly, we can expand the in-$\psi$ mode in terms of the out-$\psi$ mode as
$\psi_{k,\lambda}^{({\rm in})} =
\alpha_{k,\lambda} \psi_{k,\lambda}^{({\rm out})} + \beta_{k,\lambda} \psi_{k,\lambda}^{({\rm out}) *}$,
where we used the fact that $\langle \psi_{k,\lambda}^{({\rm in})} | \psi_{k,\lambda}^{({\rm in}) *} \rangle =
\langle \psi_{k,\lambda}^{({\rm out})} | \psi_{k,\lambda}^{({\rm out}) *} \rangle = 1$.
Consequently, we have
$A_{k,\lambda}^{({\rm in})} =
\alpha_{k,\lambda} A_{k,\lambda}^{({\rm out})} + \beta_{k,\lambda} A_{k,\lambda}^{({\rm out}) *}$,
which is the wanted relation between the electromagnetic in- and out modes.
A relation of this type is know as Bogolubov
transformation and the quantities $\alpha_{k,\lambda}$ and $\beta_{k,\lambda}$
are called Bogolubov coefficients. They satisfy the relation
$|\alpha_{k,\lambda}|^2 - |\beta_{k,\lambda}|^2  = 1$, which can be easily derived
from their defining equations.
To find the relation between the in- and out-creation and annihilation operators,
we insert the Bogolubov transformation in Eq.~(\ref{A})
and compare the result with the the expression of $\A$ defined in the out-Fock space. We find
$a_{\kk,\lambda}^{({\rm out})} = \alpha_{k,\lambda} \, a_{\kk,\lambda}^{({\rm in})}
- \beta_{k,\lambda}^* \, a_{-\kk,\lambda}^{({\rm in}) \dag}$.
From the above equation,
it follows immediately that the two Fock spaces
based on the two choices $|0, {\rm in} \rangle$ and $|0, {\rm out} \rangle$ of
the vacuum are generally different.
In particular, the in-vacuum state will contain out-particles so long as $\beta_{k,\lambda} \neq 0$:
$n_{\kk,\lambda} = \langle 0, {\rm in} |N^{({\rm out})}_{\kk,\lambda}| 0, {\rm in} \rangle = |\beta_{k,\lambda}|^2$,
where $N^{({\rm out})}_{\kk,\lambda} = a_{\kk,\lambda}^{({\rm out}) \dag} a_{\kk,\lambda}^{({\rm out})}$
is the number operator in the out-Fock space.
This situation is understood
as the creation of particles out from the vacuum by the changing expansion parameter,
and then by the changing gravitational field.
(Note that particles are created in pairs
with opposite momenta, $n_{-\kk,\lambda} = n_{\kk,\lambda}$.)
However, such a particle creation is forbidden, for symmetry reasons, in
conformal-invariant theories, such as standard electromagnetism in a Friedmann-Robertson-Walker
spacetime, a result known as ``Parker theorem''~\cite{Birrell-Davies}.
Accordingly, we must have no particle production if $f$ is a constant.

{\it Actual magnetic field.} -- What we observe today (in the out-region) is the in-vacuum expectation
value (VEV) of the squared magnetic field operator.
The latter is
\begin{equation}
\label{B}
\langle 0, {\rm in} |\B^2(\eta,\x)| 0, {\rm in} \rangle =
\sum_{\lambda=1}^2 \! \int_0^{\infty} \! \frac{dk}{k} \, {\mathcal P}_{k,\lambda}^{(\rm in)}(\eta), \nonumber
\end{equation}
where $a^2 \B = \nabla \times \A$ is the magnetic field operator,
${\mathcal P}_{k,\lambda}^{(\rm in)} = \frac12 \, {\mathcal P}_k^{\rm M} |A_{k,\lambda}^{({\rm in})}|^2$
is the magnetic power spectrum, and ${\mathcal P}_k^{\rm M}(\eta) = k^4/2\pi^2 a^4$ the
corresponding spectrum in the free Maxwell theory.
Inserting the Bogolubov transformation in ${\mathcal P}_{k,\lambda}^{(\rm in)}$, we find
\begin{equation}
\label{P}
{\mathcal P}_{k,\lambda}^{(\rm in)}(\eta) = {\mathcal P}_{k,\lambda}^{({\rm out},n)}(\eta)
+ {\mathcal P}_{k,\lambda}^{({\rm out},0)}(\eta) + {\mathcal P}_{k,\lambda}^{(\rm mix)}(\eta),
\end{equation}
where~
\begin{eqnarray}
&& \!\!\!\!\!\!\! {\mathcal P}_{k,\lambda}^{({\rm out},n)} =
                      {\mathcal P}_{k,\lambda}^{({\rm out},0)}(n_{\kk,\lambda} + n_{-\kk,\lambda}), \nonumber \\
&& \!\!\!\!\!\!\! {\mathcal P}_{k,\lambda}^{({\rm out},0)} =
                      \frac12 \, {\mathcal P}_k^{\rm M} |A_{k,\lambda}^{({\rm out})}|^2, \nonumber \\
&& \!\!\!\!\!\!\! {\mathcal P}_{k,\lambda}^{(\rm mix)} =
                      2{\mathcal P}_{k,\lambda}^{({\rm out},0)} \sqrt{n_{\kk,\lambda} (n_{\kk,\lambda} +1)}
                      \cos (\Omega_{k,\lambda} + 2 \Theta_{k,\lambda}), \nonumber
\end{eqnarray}
with $\Omega_{k,\lambda} = \mbox{Arg} (\alpha_{k,\lambda} \beta^*_{k,\lambda})$ and
$\Theta_{k,\lambda}(\eta) = \mbox{Arg} \, A_{k,\lambda}^{({\rm out})}$.
The term $\sum_{\lambda}\int \! dk k^{-1} \! \left({\mathcal P}_{k,\lambda}^{({\rm out},n)}
+ {\mathcal P}_{k,\lambda}^{({\rm out},0)}\right)$
is formally equal to
$\langle n_{\kk,\lambda}, n_{-\kk,\lambda}, {\rm out} |\B^2(\eta,\x)| n_{\kk,\lambda}, n_{-\kk,\lambda}, {\rm out} \rangle$,
where $|n_{\kk,\lambda}, n_{-\kk,\lambda}, {\rm out} \rangle$ is the state containing $n_{\kk,\lambda}$ out-particles with
wavenumber $\kk$ and helicity $\lambda$, and $n_{-\kk,\lambda}$ out-particles with wavenumber $-\kk$ and same helicity $\lambda$, so that
$\langle n_{\kk,\lambda}, n_{-\kk,\lambda}, {\rm out} |N^{({\rm out})}_{\kk,\lambda}| n_{\kk,\lambda}, n_{-\kk,\lambda}, {\rm out} \rangle =
n_{\kk,\lambda} + n_{-\kk,\lambda}$.
In particular, the term $\sum_{\lambda}\int \! dk k^{-1} {\mathcal P}^{({\rm out},0)}_{k,\lambda}$ can be
view as representing the out-vacuum magnetic fluctuations, since it is formally equal to
$\langle 0, {\rm out} |\B^2(\eta,\x)| 0, {\rm out} \rangle$.
The quantity $\sum_{\lambda}\int \! dk k^{-1} {\mathcal P}^{(\rm mix)}_{k,\lambda}$ comes from
the interference between the $\alpha_{k,\lambda} A_{k,\lambda}^{({\rm out})}$ and $\beta_{k,\lambda} A_{k,\lambda}^{({\rm out}) *}$
wave functions and is, in general, different from zero.
The physical origin of this term resides in the fact that the in-vacuum state 
is not generally proportional to the state $| n_{\kk,\lambda}, n_{-\kk,\lambda}, {\rm out} \rangle$,
in which case the interference term would be zero.

In the out-region, where $A_{k}^{({\rm out})} = e^{-ik\eta}$, we get, using Eq.~(\ref{P}),
\begin{equation}
\label{Pn}
{\mathcal P}_{k}^{(\rm in)} \! = {\mathcal P}_k^{\rm M}\!\!
\left[\frac12 + n_\kk + \sqrt{n_\kk (n_\kk +1)} \cos (\Omega_k - 2k\eta)\right]
\end{equation}
(hereafter, we drop the subscript $\lambda$ labeling photon polarization states, if not strictly necessary).

Let us concentrate on electromagnetic modes that behaves classically in the
out-region, namely modes with large occupation number, $n_\kk \gg 1$.
This is because only these modes may eventually explain the presence of cosmic magnetic fields today.
So far, we have neglected external electric currents $j^\mu$
since they are produced only after inflation,
during the process of reheating. Such currents, since generated by microphysical processes, are still vanishing
on superhorizon scales after reheating due to causality~\cite{Barrow}. Inside the horizon, instead,
and if the electromagnetic field is classical,
they can be written as $j_\mu = (0,-a \sigma_c \, \E)$, where
$a^2 \E = -\dot{\A}$ is the electric field and $\sigma_c$ is the conductivity of the primeval plasma.
The latter is very high in the early Universe, so that
post-inflationary electric currents
wash out any electric field and freeze the inflation-produced magnetic field,
which evolves then adiabatically, ${\mathcal P}_{k}^{(\rm in)} \propto a^{-4}$,
from the end of reheating till today (we are neglecting, here, any possible effect
of magnetohydrodynamic turbulence~\cite{MHD}).
Accordingly, Eq.~(\ref{Pn})
is valid up to the (positive) time $\eta_{\downarrow}$,
when a mode $k$ crosses inside the horizon after inflation. Such a time is roughly defined by
$k\eta_{\downarrow} \simeq 1$. For $\eta > \eta_{\downarrow}$, the evolution of the magnetic field
is adiabatic. From these considerations, and defining the magnetic field on a scale $1/k$ as
$B_k(\eta) = \sqrt{\sum_{\lambda}{\mathcal P}_{k,\lambda}^{(\rm in)}}$, we find, in the limit $n_\kk \gg 1$,
\begin{equation}
\label{B0}
B_k(\eta_0) \simeq \zeta_k k^2 \sqrt{n_\kk} \, ,
\end{equation}
where $\zeta_k = \sqrt{1 + \cos (\Omega_k - 2k\eta_{\downarrow})}/\pi$.
Below, for the Ratra power-law coupling, we will find that $\Omega_k$ approaches either the constant value
$\pi$ or $-\pi$ in the classical limit $n_\kk \gg 1$.
Consequently, $\zeta_k$ is an order-one constant.

The result expressed by Eq.~(\ref{B0}) is very different from the one
we find in the standard literature of inflation-produced magnetic fields~\cite{Subramanian}.
In fact, the usual reasoning, which has been recently criticized by Tsagas~\cite{Tsagas},
is the following. One evolves superhorizon magnetic modes from the beginning
up to the end of inflation. After reheating is completed, let us say at the time $\eta_e$,
such modes evolves adiabatically up to today, so that
$B_k(\eta_0) \simeq a^2(\eta_e) B_k(\eta_e)$.
This result is incorrect since it reposes on the physically incorrect assumption that magnetic modes are
frozen into the plasma even if they live on superhorizon scales.

{\it Ratra power-law coupling.} -- In the light of the above results, let us now reconsider the
generation of magnetic fields at inflation in the well-known and well-studied case
of a power-law coupling function which arises in the Ratra model.
Let us take then $f(\eta) = f_* (\eta_*/\eta)^{2p}$ if $\eta \leq \eta_e$,
and $f(\eta) = 1$ otherwise, where $\eta_e$ is the time when inflation ends.
Here, $f_* \geq 1$, $p$ is a real parameter, $\eta_* = \eta_e$ if $p<0$, and
$\eta_* = \eta_i$ if $p>0$, where $\eta_i$ is the time when inflation begins.
The Bunch-Davies-normalized $\psi_k^{({\rm in})}$ and $\psi_k^{({\rm out})}$
wave functions are easily found. They are
\begin{eqnarray}
\label{2}
\!\!\!\!\!\!\!\!\!\!\! \psi_k^{({\rm in})}(\eta) \!\!& = &\!\!
\left\{ \begin{array}{lll}
  \frac{1}{\sqrt{2k}} \, h_{\nu_p} (x), & ~~~~~~~~~~ \eta \leq \eta_e, \nonumber \\
  \frac{c_{k,1}^{(\rm in)}}{\sqrt{2k}} \, e^{ix} + \frac{c_{k,2}^{(\rm in)}}{\sqrt{2k}} \, e^{-ix},  & ~~~~~~~~~~ \eta > \eta_e,
  \end{array}
  \right.
\\
\label{3}
\!\!\!\!\!\!\!\!\!\!\! \psi_k^{({\rm out})}(\eta) \!\!& = &\!\!
\left\{ \begin{array}{lll}
   \frac{c_{k,1}^{(\rm out)}}{\sqrt{2k}} \, h_{\nu_p} (x)
   + \frac{c_{k,2}^{(\rm out)}}{\sqrt{2k}} \, h_{\nu_p}^* (x), & \eta \leq \eta_e, \nonumber \\
   \frac{1}{\sqrt{2k}} \, e^{ix},  & \eta > \eta_e. \nonumber
  \end{array}
  \right.
\end{eqnarray}
Here, $h_{\nu_p}(x) = \sqrt{\pi/2} \, e^{i \frac{\pi}{2} \left( \frac12 + \nu_p \right)} \sqrt{x} \, H_{\nu_p}^{(1)} (x)$,
$H_\nu^{(1)} (x)$ is the Hankel function of the first kind,
$\nu_p = |\frac12 + p|$, $x = -k\eta$, and $c_{k,1,2}^{(\rm in,out)}$ are constant of integrations.
The latter can be found by imposing the continuity of $\psi_k^{({\rm in})}$ and $\psi_k^{({\rm out})}$,
and their derivatives, at the time $\eta_e$. We find
$c_{k,1}^{(\rm in)} = c_{k,1}^{(\rm out) *} =
\langle \psi_k^{({\rm in})} | \psi_k^{({\rm out}) *} \rangle_{|\eta=\eta_e}  = \alpha_k$
and
$c_{k,2}^{(\rm in)} = -c_{k,2}^{(\rm out)} =
-\langle \psi_k^{({\rm in})} | \psi_k^{({\rm out})} \rangle_{|\eta=\eta_e} = \beta_k$.
The particle number is then
\begin{equation}
\label{nk}
n_\kk = \frac14 \left| \left[1 + \frac{i}{x_e}\left(\frac12 - \nu_p \right) \!\right] \! h_{\nu_p}(x_e) - h_{\nu_p-1}(x_e) \right|^{\, 2} \!,
\end{equation}
where $x_e = -k\eta_e$, and it is zero for $p=0$ (constant coupling function).
Modes that are well inside the horizon at the end of inflation, $x_e \gg 1$, are not efficiently produced,
$n_\kk \sim x_e^{-4}$ (hereafter, we neglect numerical factor in asymptotic expansions).
The opposite is true for superhorizon
modes ($x_e \ll 1$), $n_\kk \sim x_e^{-1-2\nu_p}$.
Moreover, we find $\Omega_k = \pi \, \mbox{sgn}(2\nu_p-1) + \mathcal{O}(x_e)$ for $x_e \ll 1$.
The most interesting case for inflationary magnetogenesis is when $\nu_p = 3/2$ ($p=-2$ or $p=1$)
[Eq.~(\ref{nk}) reduces to $n_\kk = 1/4x_e^4$].
This case, in fact, corresponds to a scale-invariant magnetic field $B_k$ in the out-region.
Moreover, for a scale of inflation $M$ of about $10^{16}\GeV$
(and assuming, for simplicity, an instantaneous reheating), the intensity of such a field
today, $B_k(\eta_0) \simeq 5 \times 10^{-13} \G$,
has the right value to directly explain cosmic magnetization (see, e.g.,~\cite{Campanelli2}).

{\it Backreaction and renormalization.} -- The above analysis is valid only if the
created (electro)magnetic field does not appreciably back-react on the dynamics
of the Universe.
After inflation, electric fields inside the horizon are washed out by the high conductivity of the primeval plasma,
while the sub-horizon inflation-produced magnetic field evolves adiabatically.
As it is easy to see, if such a field has to explain the observed cosmic magnetic fields,
then its energy is always subdominant with respect to that of the Universe.
During inflation, the in-VEV of the electromagnetic energy has to be much lower than the energy
density of inflation, $\langle 0,\rm in| \rho_{\rm em}(\eta,\x)|0,\rm in\rangle \ll \rho_{\rm inf}$.
Assuming, for the sake of simplicity, a de Sitter inflation, we have
$\rho_{\rm inf} = M^4 = 3H^2/(8\pi G)$, where
$H \ll M$ is the Hubble parameter during inflation
and $G$ the Newton constant. The electromagnetic energy density is
$\rho_{\rm em} = f (\E^2 + \B^2)/2$.
Using Eq.~(\ref{A}), we find
$\langle 0,{\rm in}| \rho_{\rm em}|0,{\rm in}\rangle = \int_0^{\infty} \! dk k^{-1} \rho^{({\rm in})}_k(\eta)$,
where the electromagnetic energy spectrum is
\begin{equation}
\label{E}
\rho^{({\rm in})}_{k} = \rho^{\rm M}_k \! \left[\,|h_{\nu_p}(x)|^2 + |h_{\pm \nu_{-p}}(x)|^2\right] \!,
\end{equation}
and ``$\pm$'' correspond to $|p| \geq 1/2$ and $|p| < 1/2$, respectively.
(For the scaling-invariant case, $p=-2$ or $p=1$, Eq.~(\ref{E}) reduces to
$\rho^{({\rm in})}_{k} = \rho^{\rm M}_k [1 + (1-\frac{p}{2}) x^{-2} + \frac32 (1-p) x^{-4}]$.)
For large wavenumbers, $x \gg 1$, $\rho^{({\rm in})}_{k}$ reduces to
the electromagnetic energy spectrum in the free Maxwell theory, $\rho^{\rm M}_k = x^4 H^4/2\pi^2$,
which is ultraviolet divergent. This is a typical outcome of quantum theory in
curved spacetime which can be cured by renormalization.
For superhorizon modes, $x \ll 1$, we have, instead,
$\rho^{({\rm in})}_{k} \sim x^{5-2\nu_{|p|}} H^4$.
Different schemes of renormalization exist in the literature. However, whatever scheme is adopted,
it must not change, on physical grounds, the expression of the energy spectrum in the classical limit.
Classicalization is indeed realized for superhorizon modes during inflation (see below).
Therefore, we must have that
$\rho^{({\rm in})}_{k, \rm phys} = \rho^{({\rm in})}_k$ for $x \rightarrow 0$,
where $\rho^{({\rm in})}_{k, \rm phys}$ is the renormalized (i.e. physical)
energy spectrum, and $\rho^{({\rm in})}_{k, \rm phys} = 0$ for $x \rightarrow +\infty$
in order to have a finite in-VEV of the electromagnetic energy density.
The above behaviors for $\rho^{({\rm in})}_{k, \rm phys}$ are confirmed, for example,
in the ``adiabatic renormalization'' scheme~\cite{Campanelli-Marrone}.
We conclude that backreaction on inflation is completely negligible for $|p| \leq 2$.
This leaves the scale-invariant case free from the backreaction problem.

{\it Classicalization.} -- Let us now work in the Heisenberg representation.
Following~\cite{Starobinsky,Giovannini}, we introduce the time-dependent annihilation operator
\begin{equation}
b_{\kk,\lambda}(\eta) =
\int \! d^3x \, \frac{\ee^*_{\kk,\lambda}}{\sqrt{2k}} \cdot
[k \, \boldsymbol{\Psi}(\eta,\x) +i\boldsymbol{\Pi}(\eta,\x)] \, e^{-i\kk \x}. \nonumber
\end{equation}
The re-scaled electromagnetic field $\boldsymbol{\Psi} = \sqrt{f} \, \A$ and its conjugate momentum
$\boldsymbol{\Pi} = \dot{\boldsymbol{\Psi}} -\frac12 (\dot{f}/f) \boldsymbol{\Psi}$
(and, in turns, $b_{\kk,\lambda}$ and $b_{\kk,\lambda}^{\dag}$),
satisfy the usual equal-time commutation relations.
The evolution of the annihilation and creation operators follow from the Heisenberg
equation of motions,
whose general solution can be represented through a Bogolubov transformation as
\begin{eqnarray}
&& b_{\kk,\lambda}(\eta) = \mu_{k} b_{\kk,\lambda}(-\infty) + \nu_{k} b^{\dag}_{-\kk,\lambda}(-\infty), \nonumber \\
&&b^{\dag}_{-\kk,\lambda}(\eta)  = \mu_{k}^* b^{\dag}_{-\kk,\lambda}(-\infty) + \nu_{k}^* b_{\kk,\lambda}(-\infty). \nonumber
\end{eqnarray}
The Bogolubov coefficients $\mu_{k}(\eta)$ and $\nu_{k}(\eta)$
satisfy the Bogolubov relation $|\mu_{k}|^2 - |\nu_{k}|^2 = 1$.
We can relate $\mu_{k}$ and $\nu_{k}$
to $A_k^{({\rm in})}$. Since
$b_{\kk,\lambda}(-\infty) = a_{\kk,\lambda}^{({\rm in})}$ and
$b^{\dag}_{-\kk,\lambda}(-\infty) = a_{-\kk,\lambda}^{({\rm in}) \dag}$,
a straightforward calculation gives
$\mu_{k} - \nu_{k}^* = \sqrt{f} \, A_k^{({\rm in})}$ and
$\mu_{k} + \nu_{k}^* = i \sqrt{f} \, \dot{A}_k^{({\rm in})}/k$.
We can now use the standard parametrization of the Bogolubov coefficients
in term of the squeezing parameter $r_k(\eta)$ (which, without loss of generality, we can assume
to be positive-defined), the squeezing angle $\varphi_k(\eta)$, and the phase $\theta_k(\eta)$:
$\mu_{k} = e^{-i\theta_k} \cosh r_k$ and $\nu_{k,} = e^{i(\theta_k +2\varphi_k)} \sinh r_k$.
In this case, using the expressions that relate $\mu_{k}$ and $\nu_{k}$
to $A_k^{({\rm in})}$, we find
$\rho^{({\rm in})}_k  = \rho^{\rm M}_k \cosh \! 2r_k$.

We want now to determine when the inflation-produced electromagnetic field classicalizes.
To this end, we re-consider the Bogolubov relation in physical units, $|\mu_k|^2 - |\nu_k|^2 = \hbar$,
where $\hbar$ is the Planck constant.
In the classical limit $\hbar \rightarrow 0$, the above relation is satisfied
only for $r_k \rightarrow +\infty$.
Therefore, large values of the squeezing parameter indicate that the system is in classical regime,
and we conclude that electromagnetic fields during inflation behave classically
if their energy density is much greater than the energy density in the free Maxwell theory.
For the power-law case discussed above, classicalization is then realized, as anticipated,
for superhorizon electromagnetic modes.

{\it Smooth coupling.} -- In real models of inflationary magnetogenesis, both the
expansion parameter and the coupling $f(\eta)$ are smooth ($C^\infty$) functions
of the conformal time. Moreover, in the in- and out-regions
($\eta \rightarrow \mp \infty$), the spacetime
is slowly varying in standard Friedmann universes,
in the sense that $\lim_{\eta \rightarrow \pm \infty} \frac{d^n}{d\eta^n} \, \frac{\dot{a}}{a} = 0$
for all integers $n$. If also the coupling function satisfies this last property
(and this is a very reasonable assumption),
then the in- and out-vacuum states are vacua of infinite
adiabatic order~\cite{Birrell-Davies}. This implies that the number of
produced particle goes to zero exponentially (more precisely, faster than any power of
$k$) in the limit $k \rightarrow +\infty$~\cite{Birrell-Davies}.

A numerical analysis~\cite{Campanelli-Marrone} performed by using coupling functions
which smoothly interpolate between $f(\eta) \propto \eta^{-2p}$ during inflation
and the asymptotic value $f(\eta) = 1$ in the out-region,
confirms all the results obtained in the discontinuous power-law case in the long wavelength limit
(this is because long wavelength modes are not strongly
affected by variations of the coupling function on small timescales).
Moreover, it clearly shows an exponential decay of $n_\kk$ for $k \rightarrow +\infty$,
instead of the $k^{-4}$ law derived in the discontinuous case.
(Such kind of decays can be explained analytically in semiclassical approximation~\cite{Campanelli-Marrone}.)

{\it Tensor and scalar curvature perturbations.} -- Before concluding, we mention the fact that,
in the Ratra power-law model and when backreaction on inflation is negligible, 
the electromagnetic contributions to the primordial
tensor curvature perturbations and to the ``power spectrum'', ``bispectrum'', and ``trispectrum''
of primordial scalar curvature perturbations are fully compatible with recent comic microwave
background observations~\cite{Campanelli-Marrone}.

{\it Conclusions.} -- We have pointed out that previous calculations of the actual,
inflation-produced magnetic power spectrum,
which rule out the Ratra inflaton-photon kinematic coupling model, are incorrect.
This is because post-inflationary currents, which are supposed to freeze superhorizon magnetic fields after inflation,
are indeed vanishing on superhorizon scales due to causality arguments. The correct picture is, instead, that inflationary fields
remain frozen into the primeval plasma only after they re-enter the horizon after reheating.
Relating the magnetic spectrum after inflation to the occupation number of
photons created out of the vacuum during inflation,
we have indicated the physical correct way to evolve magnetic fields
from the beginning of inflation up to today, through their re-entering the horizon.
We have found that the scale-invariant magnetic field produced in the Ratra model
directly accounts for cosmic magnetic fields if the scale of
inflation is of order of $10^{16} \GeV$. Moreover, such a field
does not back-react on inflation and it does not suffer from strong-coupling problems.

We thank A. Marrone for useful discussions.


\end{document}